\documentclass{article}
\usepackage{graphicx}
\begin{document}
\def\tect{$^{130}$Te }
\def\xect{$^{130}$Xe }
\def\tecv{$^{128}$Te }
\def\tecvt{$^{123}$Te }
\def\pbdd{$^{210}$Pb }
\def\udt{$^{238}$U }
\def\thdt{$^{232}$Th }
\def\tld{$^{208}$Tl }
\def\kq{$^{40}$K }
\def\cs{$^{60}$Co }
\def\tectn{$^{130}$Te}
\def\tecvn{$^{128}$Te}
\def\thdtn{$^{232}$Th}
\def\xectn{$^{130}$Xe}
\def\numass{m$_{\nu}$~}
\def\avl{$\langle \lambda \rangle$~}
\def\ave{$\langle \eta \rangle$~}
\def\avm{$\langle g_{\chi\nu} \rangle$~}
\def\mee{$\langle m_{ee} \rangle$~}
\def\mnu{$\langle m_{\nu} \rangle$~}
\def\amnu{$\vert \langle m_{\nu} \rangle \vert$~}
\def\mmod{$\vert \langle m_{ee} \rangle \vert$}
\def\mb{$\langle m_{\beta} \rangle$~}
\def\BBz{$\beta\beta(0\nu)$~}
\def\BBm{$\beta\beta(0\nu,\chi)$~}
\def\BBd{$\beta\beta(2\nu)$~}
\def\BB{$\beta\beta$~}
\def\Mz{$|M_{0\nu}|$~}
\def\Md{$|M_{2\nu}|$~}
\def\Tz{$T^{0\nu}_{1/2}$~}
\def\Td{$T^{2\nu}_{1/2}$~}
\def\Tm{$T^{0\nu\,\chi}_{1/2}$~}
\def\ca{$\sim$}
\def\dca{$\approx$}
\def\dot{$\cdot$}
\def\pom{$\pm$ }
\def\gm{$\gamma$}
\def\ne{$\neq$}
\def\teod{TeO$_2$~}
\def\teodn{TeO$_2$}
\def\be{\begin{equation}}
\def\ee{\end{equation}}
\def\gohm{G$\Omega$}
\def\ohm{$\Omega$}
\def\per{$\times$}
\def\ciccio{5\per5\per5 cm$^3$}
\def\magro{3\per3\per6 cm$^3$}

\title{Neutrino masses and Neutrinoless Double Beta Decay:
Status and expectations}
 
\author{Oliviero Cremonesi\thanks{e-mail: oliviero.cremonesi@mib.infn.it}
\vspace{1pc}
	\\
	INFN Sez. Milano Bicocca, Milano, Italy}

\maketitle 

\begin{abstract}
Two most outstanding questions are puzzling the world of neutrino Physics: the possible Majorana nature of neutrinos and their absolute mass scale. Direct neutrino mass measurements and neutrinoless double beta decay (\BBz) are the present strategy to solve the puzzle. Neutrinoless double beta decay violates lepton number by two units and can occurr only if neutrinos are massive Majorana particles. A positive observation would therefore necessarily imply a new regime of physics beyond the standard model, providing fundamental information on the nature of the neutrinos and on their absolute mass scale. After the observation of neutrino oscillations and given the present knowledge of neutrino masses and mixing parameters, a possibility to observe $\beta\beta(0\nu)$ at a neutrino mass scale in the range 10-50 meV could actually exist. This is a real challenge faced by a number of new proposed projects. Present status and future perpectives of neutrinoless double-beta decay ($\beta\beta(0\nu)$) experimental searches is reviewed. The most important parameters contributing to the experimental sensitivity are outlined. A short discussion on nuclear matrix element calculations is also given. Complementary measurements to assess the absolute neutrino mass scale (cosmology and single beta decays) are also discussed.
\end{abstract}
 
\section{Introduction}
\label{intro}
Neutrino oscillation experiments have provided, through the observation of the neutrino mixing and the indirect evidence for finite neutrino masses, the strongest demonstration that the Standard Model of electroweak interactions is incomplete and that new Physics beyond it must exist. Moreover, neutrinos still continue to play a key role for our understanding of the fundamental laws of Physics since some of the most relevant open questions (e.g. matter-antimatter asymmetry) seem to point towards the importance of neutrino properties. In this respect, two most outstanding questions are puzzling the world of neutrino Physics: the possible Majorana nature of neutrinos and their absolute mass scale.
In fact, the values for the squared mass differences between the neutrino species measured so far, leave room for two possibile hierarchical mass arrangements of neutrino masses (Direct and Inverted), besides the obvious quasi-degenerate option.
The oscillation experiments cannot answer at all the first question and are trying to exploit matter effects for approaching the second one. Here the problem is represented by the smallness of the neutrino masses. Present techniques for direct measurements of the electron antineutrino mass guarantee a model-independent approach but can only probe the quasi-degenerate region. On the other hand, the much more sensitive consmological inferences and neutrinoless double-beta decay experiments could sound the inverted hierarchy but suffer from a heavy model dependance. A common effort is therefore compulsory.

The paucity of neutrino masses is also related to the question of the nature of neutrino. In fact, neutrinos are the only fermions for which the Majorana formulation~\cite{major37} is possible (assuming a violation of the Lepton Number), but this description tends to be indistinguishable from the Dirac one, in the limit of vanishing masses.
Until the discovery of the massive nature of neutrinos little attention was therefore dedicated to the issue of Majorana neutrinos. On the other hand the situation has severely changed after 1998 and there is a common consensus that Majorana description is indeed the best description we can find for the physical neutrinos.
If neutrinos are Majorana particles, then there is one (and the only one experimentally viable) process that can test this property: neutrinoless double beta decay (\BBz).

The relevance of the questions of neutrino nature and absolute mass scale for our understanding of the fundamental laws of Physics has been recognized since a long time by the international scientific community and included in the world-wide strategies for the future of Astroparticle Physics~\cite{matrix,aspera}.

\section{Neutrino mass measurements}
The question of the absolute mass scale of neutrinos is presently an important issue both for particle physics and cosmology.
Direct informations on the neutrino mass can be obtained in a precise and model-independent way through a kinematical analysis of the final region of the beta decay spectra. The measured parameter is 
\begin{equation}
m_e^2=\sum_i \vert U_{ei}^2 \vert ^2 m_i^2
\end{equation}
So far, the study of the $^3$H beta decay end-point by means of electrostatic spectrometers has proved  to be the most effective, yielding and upper limit of the electron anti-neutrino mass of 2.2~eV\cite{kraus05}. The ultimate evolution of this technique is KATRIN, a huge electrostatic spectrometer characterized by a sensitivity of 0.2~eV, presently under construction at Muenster and expected to start data taking in 2012~\cite{katrin}. While further improvements look unlikely, the spectrometric approach is not free from systematic uncertainties: the measured electron energy has to be corrected for the energy lost in exciting atomic and molecular states, in crossing the source, in scattering through the spectrometer, etc. 
A possible way out to both limitations has been identified since a log time in the calorimetric approach whose only limitation seems the forced detection of all the beta decays. Since the useful fraction is given approximately by ($\Delta$E/Q)$^3$ beta decaying isotopes with the lowest Q value have been selected. $^{187}$Re has been considered for its low Q-value (\ca 2.5~keV) and because it can be used as a bolometric detector\cite{vitale}. Two experiments have been so far carried out MANU and MIBETA which have yielded limits of 26 eV and 15 eV at 95 and 90 \% C.L. respectively. Further developments in the calorimetric technique claim for sensitivities in the sub-eV region or better through the use of huge arrays of detectors. MARE~\cite{mare} is the corresponding long term project carrying the expectations for the future direct neutrino mass measurements. Originally based on the development of fast Re bolometers, MARE is now including also a $^{163}$Ho option originally, suggested by the Genova group as a unique opportunity for a self-calibrated, high statistics experiment exploiting the enhancement in sensitivity due to the closeness of the $^{163}$Ho EC Q value and the Dy atomic M lines~\cite{olmio}.

Neutrino mass is also one of the most important targets in cosmology. Since neutrino mass affects the evolution of the Universe in some observable ways, a mass constraint can be obtained from the cosmological data such as cosmic microwave background (CMB), galaxy clustering, Lyman-$\alpha$ forest, and weak lensing data. All available data sets are usually combined to obtain constraints on the sum of the neutrino mass species S$_C$=$\sum m_{\nu}$. Limits as stringent as 0.2~eV on S$_C$ can thus be obtained. They suffer however from strong dependencies from the used models which tend to spoil their reliability and the comparison with the terrestrial measurrements. Important improvements helping to fix the model uncertainties are expected as the technology will evolve providing measurements of increasing quality.

\section{Neutrinoless Double Beta Decay}
\label{NDBD}
First suggested by M.Goeppert-Mayer in 1935, Double Beta Decay (DBD) is a rare spontaneous nuclear transition in which an initial nucleus (A,Z) decays to a member (A,Z+2) of the same isobaric multiplet with the simultaneous emission of two electrons. 
In order to avoid (or at least inhibit) the occurrence of the equivalent sequence of two single beta decays, it is generally required that both the parent and the daughter nuclei be more bound than the intermediate one. Because of the pairing term, such a condition is fulfilled in nature for a number of even-even nuclei. The decay can then proceed both to the ground state or to the first excited states of the daughter nucleus. 
Double beta transitions accompanied by positron emission or electron capture are also possible. However they are usually characterized by lower transition energies and poorer experimental sensitivities. They will not be discussed in the following. We refer to the most recent reviews on \BB for a more complete treatment on the subject\cite{reviews}.\\
Among the possible \BB modes two are of particular interest, the 2$\nu$ mode (\BBd) $^A_ZX \to ^A_{Z+2}X + 2e^- + 2\overline{\nu}$, which observes the lepton number conservation and it is allowed in the framework of the Standard Model (SM) of electro-weak interactions, and the 0$\nu$ mode (\BBz) $^A_ZX \to ^A_{Z+2}X + 2e^-$ which violates the lepton number by two units and occurs if neutrinos are their own antiparticles.
A third decay mode (\BBm) in which one or more neutral bosons $\chi$ (Majorons) are emitted
$^A_ZX \to ^A_{Z+2}X + 2e^- + N\chi$
is also often considered. The interest in this decay is mainly related to the existence of Majorons, massless Goldstone bosons that arise upon a global breakdown of B--L symmetry. 
From the point of view of Particle Physics \BBz is of course the most interesting of the \BB decay modes. 
In fact, after 70 years from its introduction by W.H.~Furry\cite{furry39},  \BBz is still one of the most powerful tools to test neutrino properties: it can exist only if neutrinos are Majorana particles and it allows then to fix important constraints on the neutrino mass scale. 

When mediated by the exchange of a light virtual neutrino, \BBz rate is expressed as
\begin{equation}
[ T_{1/2}^{0\nu}]^{-1} = 
G^{0\nu}|M^{0\nu}|^2{\vert\langle m_{\nu} \rangle\vert^2}
\end{equation}
where $G^{0\nu}$ is the (exactly calculable) phase space integral, $|M^{0\nu}|^2$ is the nuclear matrix element and \mnu is a linear combination of the neutrino masses
\begin{equation}
\langle m_{\nu} \rangle \equiv \sum_{k=1}^{3}\vert
U_{ek}^L \vert ^2m_ke^{i\phi _k }
\end{equation}
which, for small masses becomes
\begin{equation}
\langle m_{\nu} \rangle =
c_{12}^2 c_{13}^2 m_1+s_{12}^2 c_{13}^2 e^{i\alpha_1} m_2+s_{13}^2 e^{i\alpha_2}m_3
\end{equation}

Unfortunately, the presence of the Majorana phases $\alpha_k$ in the \mnu expression implies that cancellations are possible. Such cancellations are complete for a Dirac neutrino since it is equivalent to two degenerate Majorana neutrinos with opposite CP phases. This stresses once more the fact that \BBz can occur only through the exchange of Majorana neutrinos.
It should be also pointed out that \BBz represents the unique possibility to measure the neutrino Majorana phases.

Given the evidence for neutrino oscillations and the present knowledge of neutrino masses and mixing parameters, two possible orderings are possible (neutrino mass hierarchy problem). In the case that fortcoming \BBz experiments would not  observe any decay (and assuming that neutrinos are Majorana particles) the inverse ordering could finally be excluded thus fixing the problem of the neutrino absolute mass scale\cite{strum08,petco05}.

\mnu is actually the only \BBz measurable parameter containing direct information on the neutrino mass scale. Unfortunaltely its derivation from the experimental results on \BBz half-lifetimes requires a precise knowledge of the transition Nuclear Matrix Elements M$^{0\nu}$(NME). Many (often conflicting) evaluations are available in the literature. However, in many cases they have demonstrated in considerable disagreement among themselves, leading to large uncertainty ranges for \mnu. This has been recognized as a critical problem by the \BB community. 

\section{Nuclear Matrix Elements}
The calculation of \BBz nuclear matrix elements (NME) has been carried out in the last decades by many authors using mainly the Quasiparticle Random Phase Approximation (QPRPA, RQPRPA, pnQPRPA etc.) or the Shell Model.
The two methods have complementary virtues. While in fact QRPA calculations include many single-particle levels outside a relatively small inert core, they can hardly manage correlations. On the other hand, the shell model can include arbitrarily complicated correlations, but is limited to a few single-particle orbitals outside the inert core. 

\begin{figure}
      \centering\includegraphics[width=0.9\textwidth]{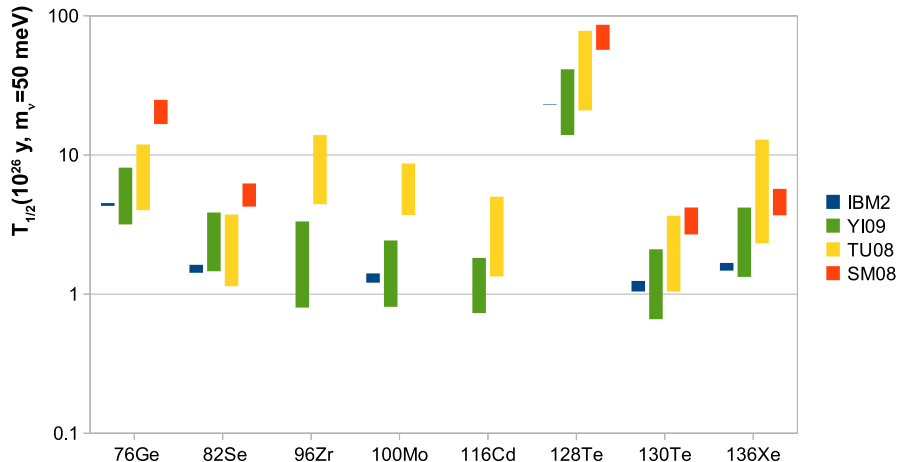}%
\caption{Expected \BBz half lives for 50 meV effective neutrino mass and different NME calculations: IBM2~\cite{ibm2}, YI09~\cite{yi09}, TU08~\cite{tu08} and SM08~\cite{sm08}.}
\label{fig:nme}

\end{figure}

Although significative improvements have been obtained recently the QRPA matrix elements still exceed those of the shell model by factors of up to about two in the lighter isotopes (e.g. $^{76}$Ge and $^{82}$Se), and somewhat less in the heavier isotopes (Fig.\ref{fig:nme}).

New calculations have been recently carried out by the group of F.~Iachello with a completely different approach based on the Interacting Boson Model (IBM2)\cite{ibm2}. The results are in reasonable agreement with QRPA matrix elements. Although these new calculations don't provide yet an answer to the question of which method is closer to the truth, they can help to identify the important effects responsible for the observed disagreement. 

\begin{figure}
      \centering\includegraphics[width=0.7\textwidth]{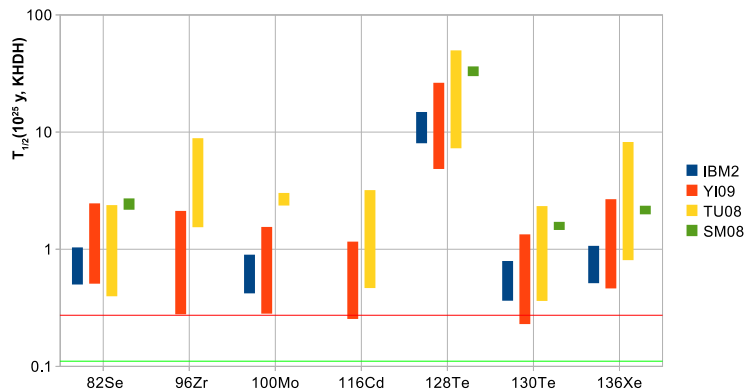}%
\caption{\BBz half life intervals corresponding to the 90\% CL range of the KHDH claim, rescaled (T$_k$=T$_{Ge}$G(Ge)M(Ge)$^2$/G(k)M(k)$^2$) according to different NME calculations: IBM2~\cite{ibm2}, YI09~\cite{yi09}, TU08~\cite{tu08} and SM08~\cite{sm08}. Available lower limits for $^{130}$Te (CUORICINO: red line) and $^{82}$Se (NEMO3: green line) are also shown.}
\label{fig:klap}
\end{figure}

The careful check of the models in order to account for the omitted physics or the important missing informations seems the only way out of the problem. A systematic analysis of the calculation methods and their basic hypotheses have been therefore started. The results should not be late to arrive although the inclusion of the missing correlations into the QRPA looks a very difficult task (because of the several uncontrolled approximations of the method) while for the shell model, at least in principle, a systematic procedure for adding the effects of missing states exists.

The ultimate limitation of the QRPA method seems the perturbative approach which is implemented in a renormalized nuclear interaction and requires always some adjustment to the data. Reasonably good results are usually obtained by a proper parametrization of the short range correlations or the reduction of the axial-vector coupling constant g$_A$. This corresponds to a phenomelogical correction of the \BBz operator whose reliability is not easy to assess.
An better approach could consist in obtaining an effective double-beta-decay operator~\cite{engel09}. 

A statystical analysis of the different NME calculation (comparison of different methods and model parameters) has also been recently considered\cite{nmestat}. Besides providing useful recipes for the comparison of the experimental results on different isotopes this approach could help in identifying systematic effects in the different calculations.

\section{\BBz experimental approaches: present status}
The only experimentally available informations in \BBz are those carried by the daughter nucleus and the two emitted electrons. Different signatures depend therefore on the number of such informations which are actually measured: sum of the electron energies, single electron energy and angular distributions, identification and/or counting of the daughter nucleus. A better signature is often synonymous of a lower background and, definitely, of a better sensitivity. All experiments tend therefore to find a compromise between the desire to collect the maximum number of informations and the best way in which such a goal can be accomplished.
Two main general approaches have been followed so far for \BB experimental investigation: i) indirect or inclusive methods, and  ii) direct or counter methods.
Inclusive methods are based on the measurement of anomalous concentrations of the daughter nuclei in properly selected samples, characterized by very long accumulation times. They include Geochemical and Radiochemical methods which, being completely insensitive to different \BB modes, can only give indirect evaluations of the \BBz and \BBd lifetimes. They have played a crucial role in \BB searches especially in the past.

Counter methods are based instead on the direct observation of the two electrons emitted in the decay. Different experimental parameters (energies, momenta, topology, etc) can then be registered according to the different capabilities of the employed detectors. These methods are further classified in {\em inhomogeneous} (when the observed electrons originate in an external sample) and {\em homogeneous} experiments (when the source  of \BB's serves also as detector). 

In most cases the various \BB modes are separated just on the base of the different distribution expected for the electron sum energies: a continuous bell ditribution for \BBd and \BBm, and a sharp line at the transition energy for \BBz. 
Direct counting experiments with very good energy resolution are presently the most attractive approach for \BBz searches.

Experimental evidence for several \BBd decays has been provided using the measured two-electron sum energy spectra, the single electron energy distributions and the event topology\ref{tab:BBres}. 
On the other hand, impressive progress has been obtained during the last years also in improving \BBz half-life limits for a number of isotopes. The best results are still maintained by the use of isotopically enriched HPGe diodes for the experimental investigation of $^{76}$Ge (Heidelberg-Moscow\cite{hmosc01} and IGEX\cite{aalse02}) but two other experiments have recently reached comparable sensitivities: NEMO3\cite{barab06,klang09} at LSM and CUORICINO at LNGS\cite{arnab08}. The former is a large inhomogeneous detector aiming at overcoming the intrinsic limits of the technique (relatively small active masses) by expanding the setup dimensions; the big advantage of the NEMO3 technique is the possibility to access single electron informations.  CUORICINO is, on the other hand, a TeO$_2$ granular calorimeter based on the bolometric technique; it aims at exploiting the excellent performance of the bolometers (and the possibility they offer to be built with any material of practical interest\cite{fiori84,alepe}) to scan the most interesting \BBz active isotopes. NEMO3 will continue data taking until the end of 2010 while CUORICINO was stopped in June 2008 to be sustituted by CUORE-0, the first tower of CUORE. The NEMO3 effort to cover as many as possible \BB nuclei thus allowing a diret check for \BBd NME elements is evident  (Tab. \ref{tab:BBres}). \\

The evidence for a \BBz signal has also been claimed (and recently confirmed \cite{klapd08} by a small subset (KHDK) of the HDM collaboration at LNGS with $T^{0\nu}_{1/2}=2.23^{+0.44}_{-0.31}\times 10^{25}$ y. The result is based on a re-analysis of the HDM data. Such a claim has raised some criticism but cannot be dismissed out of hand. On the other hand, none of the existing experiments can rule out it (fig.~\ref{fig:klap}), and the only certain way to confirm or refute it is with additional sensitive experiments. In particular, next generation experiments should easily achieve this goal.

\begin{table*}[htb]
\label{tab:BBres}
\begin{center}
\newcommand{\m}{\hphantom{$-$}}
\newcommand{\cc}[1]{\multicolumn{1}{c}{#1}}
\renewcommand{\tabcolsep}{0.9pc} 
\begin{tabular}{@{}lcclcc}
\hline
Isotope 	& T$_{1/2}^{2\nu}$            & T$_{1/2}^{0\nu}$     &  Future & Mass & Lab\\ 
 	& (10$^{19}$y)           & (10$^{24}$y)     & Experiment & (kg) & \\ 
\hline
$^{48}$Ca          & $(4.4^{+0.6}_{-0.5})$	&
	$>0.0014 $\cite{ogawa04}                & CANDLES && OTO\\
$^{76}$Ge          & $(150\pm 10)$ &
	$>19$\cite{hmosc01}                      &GERDA&18-40&LNGS\\
                   & &
	$22.3^{+4.4}_{-3.1}$\cite{klapd08}                      & &&\\
                   & &
	$>15.7$\cite{aalse02}                     &MAJORANA&60&SUSEL\\
$^{82}$Se          & $(9.2 \pm 0.7)$ &
	$>0.36$ \cite{klang09} & SuperNEMO& 100&LSM\\
$^{96}$Zr  & $(2.3\pm0.2)$ &
	$>0.0092$\cite{klang09}& &&\\
$^{100}$Mo         & $(0.71\pm0.04)$&
	$>1.1$\cite{klang09}                      & MOON&& OTO\\
$^{116}$Cd         & $(2.8\pm 0.2)$&
	$>0.17$\cite{danev03}                         & &&\\
$^{130}$Te         & $(68\pm12)$&
	$>2.94$                       & CUORE&204&LNGS\\
$^{136}$Xe         & $>81$\cite{gavri00} &
	$>0.12$\cite{berna02}                       & EXO&160&WIPP\\
		   &  &  & KAMLAND&200&KAMIOKA\\
$^{150}$Nd         & $(0.82\pm 0.09)$ &
	$>0.0036$\cite{barab05}                        &SNO+&56&SNOLAB \\
\hline
\end{tabular}
\end{center}
\end{table*}

\section{Forthcoming experiments}
The performance of the different \BBz experiments is usually expressed in terms of an experimental {\em sensitivity} or detector {\em factor of merit}, defined as the process half-life corresponding to the maximum signal n$_B$ that could be hidden by the background fluctuations at a given statistical C.L. At 1$\sigma$ level (n$_B$=$\sqrt{BTM\Delta}$), one obtains:
\begin{eqnarray}
\label{eq:sensitivity}
F_{0\nu} = \tau^{Back.Fluct.}_{1/2}=
\ln 2~N_{\beta\beta}\epsilon\frac{T}{n_B} \nonumber 
= \ln 2\times \frac{x ~\eta ~ \epsilon ~ N_A}{A} 
\sqrt{ \frac{ M ~ T }{B ~ \Delta} } ~ (68\% CL)
\end{eqnarray}  
where B is the background level per unit mass and energy, M is the detector mass, T is the measure time, $\Delta$ is the FWHM energy resolution, N$_{\beta\beta}$ is the number of \BB decaying nuclei under observation, $\eta$ their isotopic abundance, N$_A$ the Avogadro number, A the compound molecular mass, $x$ the number of \BB atoms per molecule, and $\epsilon$ the detection efficiency.\par
Despite its simplicity, equation (\ref{eq:sensitivity}) has the unique advantage of emphasizing the role of the essential experimental parameters: mass, measuring time, isotopic abundance, background level and detection efficiency. Most of the criteria to be considered when optimizing  the design of a new \BBz experiment follow directly from it: i) a well performing detector (e.g. good energy resolution and time stability) giving the maximum number of informations (e.g. electron energies and event topology); ii) a reliable and easy to operate detector technology requiring a minimum level of maintenance (long underground running times); iii) a very large (possibly isotopically enriched) mass, of the order of one ton or larger; iv) an effective background suppression strategy. Unfortunately, these simple criteria are often conflicting and simultaneous optimisation is rarely possible. 

Of particular interest is the case when the background level B is so low that the expected number of background events in the region of interest along the experiment life is of order of unity: $BMT\Delta \simeq O(1)$. In such cases one generally speaks of ''zero background'' (0B) experiments, a condition met by a number of future projects. In such conditions, eq. (\ref{eq:sensitivity}) is no more valid and the sensitivity is given by
\begin{eqnarray}
\label{eq:0sensitivity}
F_{0\nu}^{0B} = 
\ln 2~N_{\beta\beta}\epsilon\frac{T}{n_L} \nonumber 
= \ln 2\times \frac{x ~\eta ~ \epsilon ~ N_A}{A} 
\frac{ M ~ T }{n_L}
\end{eqnarray}  

where $n_L$ is a constant depending on the chosen CL and on the actual number of observed events. 
The most relevant feature of eq. (\ref{eq:0sensitivity}) is that it does not depend on the background level or the energy resolution and scales linearly with the sensitive mass M and the measure time T. Since T is usually limited to a few years and $|Delta$ is usually fixed for a given experimental technique, the 0B condition translates to $BM \simeq O(1)$. This means that for a given mass M there exists a threshold for B below which no further improvement of the sensitivity is obtained. This means that it can be useless to reduce at will the background level without a corresponding increase of the experimental mass. For typical experimental conditions $B_T \simeq \frac {1}{10~M}$ or $10~{-4}$ for a O(1t) experiment.

Analogous comments hold, on the other hand, for the discovery potential usually defined in terms of the ratio of the observed effect and background events. Also in this case, in the 0B regime the background contribution is constant and the  discovery potential scales linearly with $MT$.

Calorimetric detectors are usually preferred for future experiments since they have produced so far the best results. The calorimetric approach suffers however from a strong limitation: it can be applied only to a  small number of \BBz isotopes (e.g. $^{76}$Ge, $^{136}$Xe, $^{48}$Ca), thus limiting the number of experimentally accessible isotopes. As suggested by the NEMO3 and CUORICINO experience however, a possible way out exists. 

A series of new proposals has been boosted in recent years by the renewed interest in \BBz following neutrino oscillation results. The ultimate goal is to reach sensitivities such to allow an investigation of the inverted hierarchy (IH) of neutrino masses (\mnu\ca10-50 meV). From an experimental point of view this corresponds however to active masses of the order of 1~ton with background levels of the order of 1 c/keV/ton/y. A challenge that can hardly be faced by the current technology. Phased programs have been therefore proposed in USA and Europe\cite{matrix,aspera}.

Second generation experiments are all characterized by hundred kg detectors and 1-10 c/kev/ton background rates. Their goal is to select the best technology and approach the IH region.

A list of some of the forthcoming  \BBz projects is given in table \ref{tab:BBres}. 
They can be classified in three broad classes: i) dedicated experiments using a conventional detector technology with improved background suppression methods (e.g. GERDA, MAJORANA); ii) experiments using unconventional detector (e.g. CUORE) or background suppression (e.g. EXO, SuperNEMO) technologies; iii) experiments based on suitable modifications of an existing setup aiming at a different search (e.g. SNO+, KAMLAND). In some cases technical feasibility tests are required, but the crucial issue is still the capability of each project to pursue the expected background suppression. Although all proposed projects show interesting features for a second generation experiment, only few of them are characterized by a reasonable technical feasibility within the next few years. 

MAJORANA and GERDA are both phased programs representing large scale extensions of past successful experiments on $^{76}$Ge \BBz. 

Evolved from the HM experiment, GERDA\cite{abt04} aims at implementig the concept of Ge diodes immersed in a LAr bath\cite{klapd01} for a radical background suppression. The GERDA setup construction is presently being completed in Gran Sasso. 18 and 40 kg of Germanium detectors enriched in $^{76}$Ge are foreseen for the first and second phase respectively. GERDA-I will scrutinize the KHDK claim starting in 2010 and reaching a sensitivity T$_{1/2} > 2\times10^{25}$ y (90~\% CL) after two years of data taking. 40 kg of germanium isotopically enriched in $^{76}$Ge are already available for GERDA-II. A large part of the efforts are presently directed to develop the segmented detectors crucial for the targeted 10$^{-3}$ c/keV/kg background level. The expected 5y sensitivity is \ca $2.5\times10^{26}$ y.
Depending on the physics results of the previous phases, a third phase using 500 to 1000 kg of enriched germanium detectors is planned, merging GERDA with the US lead Majorana collaboration.

MAJORANA, a mainly USA proposal with important Canadian, Japanese, and Russian contributions, is an evolution of the IGEX experiment. The proposed initial configuration\cite{aalse01} would consist of 171 segmented n-type germanium crystals (180 kg), distributed in 3 independent ultra-clean electro-formed conventional cryostats of 57 crystals each. The whole assembly would be enclosed in a low-background passive shield and active veto and be located deep underground. A 60 kg demonstrator (single cryostat) is presently being developed to demonstrate the viability of the technique. The completion of this phase is expected in 2014.


CUORE\cite{qprop05} ({\em Cryogenic Underground Detector for Rare Events}) is a very large extension of the \teod bolometric array concept pioneered by the Milano group at the Gran Sasso Laboratory since the eighties. CUORE will consist of a rather compact cylindrical structure of 988 cubic natural \teod crystals of 5~cm side (750~g), arranged into 19 separated {\em towers} (13 {\em planes} of 4 crystals each) and operated at a temperature of ~10~mK. The expected energy resolution is \ca5~keV FWHM at the \BBz transition energy (\ca 2.53~MeV). A background level of of the order of \ca 0.01 c/keV/kg/y is expected by extrapolating the CUORICINO background results and the dedicated CUORE R\&D measurements. The expected 5y sensitivity is $2.1\times10^{26}$ y. CUORE will therefore allow a close look at the IH region of neutrino masses. CUORE is fully funded and presently under construction at LNGS at a relatively low cost thanks to the high natural abundance of \tect. Setup completion is expected in 2012.
Thanks to the bolometer's versatility, alternative options with respect to \teod are also possible. In particular, promising results have been recently obtained with scintillating bolometers\cite{pirro05} which could allow to study in the future new \BBz active isotopes with improved sensitivity.
  
EXO\cite{danil00} ({\it Enriched Xenon Observatory}) is a challenging project based on a large mass (\ca~1--10 tons) of isotopically enriched (85\% in  $^{136}$Xe) Xenon. An ingenuous tagging of the doubly charged Ba isotope produced in the decay ($^{136}Xe\to^{136}Ba^{++}+ 2e^-$) would allow an excellent background suppression. The technical feasibility of such an ambitious project aiming at a complete suppression of all the backgrounds requires a hard, still ongoing R\&D phase. The unavoidable \BBd contribution is a serious concern due to the poor energy resolution of Xe detectors. 
A smaller prototype experiment with a Xe mass of 200 kg (80\% $^{136}$Xe), is presently being installed at WIPP. The prototype has no barium tagging. The primary goal is to measure $^{136}$Xe \BBd and to study \BBz with a sensitivity of \ca 10$^{25}$ y in two years of data taking.

The proposed Super-NEMO experiment is the only based on an inhomogeneous approach. It is an extension of the successful NEMO3 concept, properly scaled  in order to accommodate \ca100 kg of $^{82}$Se foils spread among 20 detector modules. The proposed geometry is planar. The energy resolution will be improved from 12\% FWHM to 7\% FWHM to improve the signal detection efficiency from 8\% to 40\% and reduce the \BBd contribution. The detector modules will have an active water shield to further reduce cosmic ray backgrounds. The proposed detector dimensions will require a larger hall than is currently available at Frejus and an expansion of the facility is therefore required and actively pursued. A demonstrator (single module) is presently fully funded to be completed in 2011 with a test run in the current NEMO3 site. If funded, Super-NEMO construction should immediately start.  

A novel detection concept has been recently proposed by a mainly Spanish collaboration headed by the Valencia group~\cite{next}. The concept is based on a Time Projection Chamber (TPC) filled with high-pressure gaseous xenon, and with capabilities for calorimetry and tracking. Thanks to an excellent energy resolution (\ca 1\% at 2580 keV), together with a powerful background rejection provided by the distinct double-beta decay topological signature, the NEXT collaboration aims at a phased program starting with a 100 kg TPC capable of exploring the 100 meV region hence analysing the KHDH claim. Expected to operate in the Canfranc Underground Laboratory (LSC) and characterized by a projected background level of the order of 10$^{-3}$ c/keV/kg/y, NEXT-100 will be large enough  to prove the scalability of the technology up to a 1-ton detector.

New developments have been recently proposed concerning the possibility to disperse \BBz active isotopes in large masses of low-activity scintillators. SNO+ is pursuing the goal of studying $^{150}$Nd with 50 to 500 kg of isotopically enriched Neodimium depending on the results of the currently ongoing R\&D program.

A similar approach is proposed by KAMLAND, but for $^{136}$Xe. Their program should start in 2011 (first phase) with 200-400 kg of isotope and continue in 2013 with 1 ton of Xenon enriched to 90\% in $^{136}$Xe. A preliminary estimate of the 5y sensitivity for phase I amounts to \ca 10$^{26}$ y.

\section{Conclusions}
A renewed interest in the experimental study of neutrino properties has been stimulated by neutrino oscillation results. Neutrinoless \BB is finally recognized as a unique tool to measure Majorana nature of the neutrino providing in the meanwhile important informations on the neutrino mass scale and intrinsic phases, unavailable to the other neutrino experiments. 
Present \mnu sensitivities are still outside the range required to test the inverted neutrino mass hierarchy. An international effort is however supporting a phased \BBz program based on a number of newly proposed experiments to pursue such a goal.
The success of such a program strongly depends on the true capability of the proposed projects to reach the required background levels in the \BBz region. 
The claimed evidence for a \BBz signal in the HM data could be soon verified by the presently running experiments and in any case, by the forthcoming next generation experiments.\\
KATRIN will the only future experiment able to provide a model independent measurement of the electron antineutrino mass in the sub-eV region. A long term R\&D program for the development of a large calorimetric detector has been proposed to go beyond the KATRIN sensitivity and fill the gap created by the spectrometric technique limitations. Besides the development af a proper detector and the choice of the most suitable technique, a serious limitation could be provided by the huge number of required low temperature installations.\\
Increasingly stringent constraints on the sum of the neutrino masses are provided by the combined analyses of cosmological data sets. Unfortunately they still suffer from being strongly model dependent. Technological improvements are however improving the quality of the data sets and helping to fix the model parameters.\\
In conclusion, present and next future experiments will guarantee a complementary set of informations that will surely improve our knowledge of the neutrino properties. In particular, if neutrinos are Majorana fermions, next generation experiments will be able to solve the problem of the neutrino mass hierarchies.

\end{document}